\documentstyle{article}

\makeatletter
\newdimen\normalarrayskip              % skip between lines
\newdimen\minarrayskip                 % minimal skip between lines
\normalarrayskip\baselineskip
\minarrayskip\jot
\newif\ifold             \oldtrue            \def\new{\oldfalse}
\def\arraymode{\ifold\relax\else\displaystyle\fi} % mode of array entries
\def\eqnumphantom{\phantom{(\theequation)}}     % right phantom in eqnarray
\def\@arrayskip{\ifold\baselineskip\z@\lineskip\z@
     \else
     \baselineskip\minarrayskip\lineskip2\minarrayskip\fi}
\def\@arrayclassz{\ifcase \@lastchclass \@acolampacol \or
\@ampacol \or \or \or \@addamp \or
   \@acolampacol \or \@firstampfalse \@acol \fi
\edef\@preamble{\@preamble
  \ifcase \@chnum
     \hfil$\relax\arraymode\@sharp$\hfil
     \or $\relax\arraymode\@sharp$\hfil
     \or \hfil$\relax\arraymode\@sharp$\fi}}
\def\@array[#1]#2{\setbox\@arstrutbox=\hbox{\vrule
     height\arraystretch \ht\strutbox
     depth\arraystretch \dp\strutbox
     width\z@}\@mkpream{#2}\edef\@preamble{\halign
\noexpand\@halignto
\bgroup \tabskip\z@ \@arstrut \@preamble \tabskip\z@ \cr}%
\let\@startpbox\@@startpbox \let\@endpbox\@@endpbox
  \if #1t\vtop \else \if#1b\vbox \else \vcenter \fi\fi
  \bgroup \let\par\relax
  \let\@sharp##\let\protect\relax
  \@arrayskip\@preamble}
\def\eqnarray{\stepcounter{equation}%
              \let\@currentlabel=\theequation
              \global\@eqnswtrue
              \global\@eqcnt\z@
              \tabskip\@centering
              \let\\=\@eqncr
              $$%
 \halign to \displaywidth\bgroup
    \eqnumphantom\@eqnsel\hskip\@centering
    $\displaystyle \tabskip\z@ {##}$%
    \global\@eqcnt\@ne \hskip 2\arraycolsep
         $\displaystyle\arraymode{##}$\hfil
    \global\@eqcnt\tw@ \hskip 2\arraycolsep
         $\displaystyle\tabskip\z@{##}$\hfil
         \tabskip\@centering
    &{##}\tabskip\z@\cr}
\begingroup\ifx\undefined\newsymbol \else\def\input#1 {\endgroup}\fi
\input amssym.def \relax
\input amssym
\newfont{\hr}{msbm10}
\newfont{\ams}{msam10}
\font\numbers=cmss12
\font\upright=cmu10 scaled\magstep1
\def\stroke{\vrule height8pt width0.4pt depth-0.1pt}
\def\topfleck{\vrule height8pt width0.5pt depth-5.9pt}
\def\botfleck{\vrule height2pt width0.5pt depth0.1pt}
\def\Zmath{\vcenter{\hbox{\numbers\rlap{\rlap{Z}\kern 0.8pt\topfleck}\kern
2.2pt
                   \rlap Z\kern 6pt\botfleck\kern 1pt}}}
\def\Qmath{\vcenter{\hbox{\upright\rlap{\rlap{Q}\kern
                   3.8pt\stroke}\phantom{Q}}}}
\def\Nmath{\vcenter{\hbox{\upright\rlap{I}\kern 1.7pt N}}}
\def\Cmath{\vcenter{\hbox{\upright\rlap{\rlap{C}\kern
                   3.8pt\stroke}\phantom{C}}}}
\def\Rmath{\vcenter{\hbox{\upright\rlap{I}\kern 1.7pt R}}}
\def\Z{\ifmmode\Zmath\else$\Zmath$\fi}
\def\Q{\ifmmode\Qmath\else$\Qmath$\fi}
\def\N{\ifmmode\Nmath\else$\Nmath$\fi}
\def\C{\ifmmode\Cmath\else$\Cmath$\fi}
\def\R{\ifmmode\Rmath\else$\Rmath$\fi}
\def\res{{\rm res}}

\def\Im{{\rm Im}}

\def\2{{1\over 2}}
\def\N2{${\cal N}=2$}

\def\be{ \begin{eqnarray} }
\def\ee{ \end{eqnarray} }

\def\bea{\begin{eqnarray}}
\def\eea{\end{eqnarray}}

\def\beq{\begin{equation}}
\def\eeq{\end{equation}}
\def\ba{\beq\new\begin{array}{c}}
\def\ea{\end{array}\eeq}
\def\be{\ba}
\def\ee{\ea}

\renewcommand{\theequation}{\arabic{equation}}

\newcommand{\tr}{\,{\rm tr}\,}

\def\pint{\int\hspace{-1.17em}\not\hspace{0.6em}}

\def\be{\begin{equation}}
\def\ee{\end{equation}}
\def\bea{\begin{eqnarray}}
\def\eea{\end{eqnarray}}

\def\thetag#1{({\rm{\ref{#1}}})}

\title{{\bf Matrix models vs. Seiberg--Witten/Whitham theories}
\vspace{.5cm}}
\author{{\bf L.
Chekhov}\thanks{E-mail: \ chekhov@mi.ras.ru. The work is supported by
the RFBR Grant No.~02--01--00484,
the Grant for Support of the Scientific Schools 00-15-96046,
and by the Program Nonlinear Dynamics and Solitons.}
\date{ } \\ {\small
{\it Steklov Mathematical Institute, Moscow, Russia}}\\ \\
and\\ \\
{\bf A. Mironov}\thanks{E-mail: \ mironov@itep.ru; mironov@lpi.ac.ru.
The work is supported
by grants INTAS 00-334, RFBR-00-02-16477-a,
the Grant of Support for the Scientific
Schools 96-15-96798
and by the Volkswagen-Stiftung.}
\date{ } \\
{\small {\it Theory Department, Lebedev Physical Institute, Moscow, Russia}
and {\it ITEP, Moscow, Russia}}}

\begin{document}

\maketitle

\vspace{-9cm}

\begin{center}
\hfill FIAN/TD-21/02\\
\hfill ITEP/TH-40/02\\
\hfill hep-th/0209085
\end{center}

\vspace{9cm}

\begin{abstract}
We discuss the relation between matrix models and
the Seiberg--Witten type (SW) theories, recently proposed by
Dijkgraaf and Vafa. In particular, we prove that the partition
function of the Hermitean one-matrix model in the planar (large $N$)
limit coincides with the prepotential of the corresponding SW theory.
This partition
function is the logarithm of a Whitham $\tau$-function. The
corresponding Whitham hierarchy is explicitly constructed.
The double-point problem is solved.
\end{abstract}

%\eop

\paragraph{1.}

It is well known that partition functions of matrix models are
$\tau$-functions of integrable hierarchies of the Toda type \cite{GMMMO}.
In the specific double scaling limit, these $\tau$-functions become
$\tau$-functions of various reduction of the KP hierarchy \cite{FKN}.
If one makes the simplest, large-$N$ (planar) limit, the partition
function becomes the $\tau$-function of the dispersionless Toda
hierarchy, which in turn becomes
the $\tau$-function of the dispersionless (reductions of) KP
hierarchy \cite{TakTak}
after performing the continuum limit (which basically
means working nearby a singularity of the partition function).
All these dispersionless hierarchies are
just Whitham equations over trivial solutions to integrable
(Toda, KP) hierarchies.

When solving matrix models, most attention was paid to one-cut solutions
where the limiting eigenvalue distribution spans one interval on the
real axis~\cite{OC}. The results on multi-cut solutions \cite{MCU} were
few~\cite{Jur,MC,Ak}. Recently, Dijkgraaf and Vafa proposed~\cite{DV}
the new insight on the multi-cut large-$N$ limit of matrix models. Namely,
they associated this limit with a Riemann surface and
some related SW system. Its prepotential, which we prove here to be the
logarithm of the large-$N$ partition function, is typically
associated with the logarithm of some Whitham $\tau$-functions
\cite{GKMMM,RG}. This hints that the matrix matrix model in the large
$N$ limit of multi-cut type describes the Whitham system over a
non-trivial, finite-gap solution to integrable (Toda, KP) hierarchy.
In particular, this
solution passes to a finite-gap solution
of (reductions of) the KP hierarchy in the continuum limit.

In this paper, we restrict ourselves with the simplest example of the
Hermitean one-matrix model. We show that coefficients of the potential of
the model gives rise to Whitham flows and manifestly construct this Whitham
system. In fact, the authors of \cite{DV} associated
the $N=1$ SUSY gauge theory studied in \cite{CIV}
with the SW system
related to the multi-cut planar limit of matrix models.
From the point of view of $N=1$ SUSY theory,
these coefficients must be identified with couplings in the tree
superpotential, while the SW moduli are associated with v.e.v.'s of the
gluino condensates. This gives an interpretation of the results of
\cite{CIV} in the Whitham hierarchy terms.

\paragraph{2.}

We call SW system \cite{SW}\footnote{Various properties of such systems can
be found in \cite{SWbook,GM}.} the following set of data\footnote{Our
definition of the SW prepotential does not imply any connection with
prepotentials of $N=2$ SUSY gauge theories \cite{GM}. Moreover, the
prepotentials discussed in this paper are rather related to
superpotentials of $N=1$ SUSY theory \cite{CIV}.}:

\begin{itemize}

\item a family ${\cal M}$ of Riemann surfaces (complex curves) ${\cal C}$
whose dimesnion coincides with the genus;\footnote{This restriction can be
waved, see examples in \cite{WDVVlong}.}

\item a meromorphic differential $dS$ whose variations w.r.t. moduli of
curves are holomorphic.
\end{itemize}

This data allows one to define the notion of prepotential \cite{SW,IM}
related to some integrable system \cite{GKMMM}.

Indeed, one can introduce variables
\be\label{a}
a_i\equiv\oint_{A_i}dS
\ee
where $A_i$ are $A$-cycles on ${\cal C}$. Then,
\be
d\omega_i\equiv {\partial dS\over\partial a_i}
\ee
are canonical holomorphic differentials on ${\cal C}$ (normalized so that
$\oint_{A_i}d\omega_j=\delta_{ij}$). Then, introducing $B$-cycles conjugated
to $A$-cycles: $A_i\circ B_j=\delta_{ij}$, where $\circ$ means intersection,
we obtain that
\be
{\partial\over\partial a_i}\oint_{B_j}dS=\oint_{B_i}d\omega_j=T_{ij}
\ee
is the period matrix of ${\cal C}$ and is therefore symmetric. Hence,
there exists a prepotential ${\cal F}$ such that
\be
\label{F}
{\partial{\cal F}\over\partial a_i}=\oint_{B_i}dS
\ee

\paragraph{3.}

Let us consider the Hermitean one-matrix model. Its partition function is
given by the integral over Hermitean $N\times N$ matrix $M$
\be
Z_N=\int{\cal D}Me^{-N\tr V(M)}
\ee
where ${\cal D}M$ is the Haar measure on Hermitean matrices and the potential
$V(x)$ is a polynomial of degree $n+1$. After integrating out angular
variables, we obtain \cite{Mehta}
\be\label{s1}
Z_N\sim \int\prod_id\lambda_ie^{-NV(\lambda_i)+\sum_{j\ne i}\log (\lambda_i
-\lambda_j)}
\ee
In the large $N$ limit, it is standard
to introduce the density of eigenvalues
\be
\rho(\lambda)\equiv{1\over N}\sum_i\delta(\lambda-\lambda_i).
\ee
Then, (\ref{s1}) can be rewritten as
\be
Z_N\sim\int\prod_id\lambda_ie^{-N^2\left[\int\rho(\lambda)V(\lambda)
d\lambda-\int\!\int\rho(\lambda)\rho(\lambda')
\log(\lambda-\lambda')d\lambda d\lambda'
\right]}
\ee
In the large $N$ limit, this integral
can be evaluated by the saddle point method. We then
assume $\rho(\lambda)$ to be a continuous function
such that
\be\label{s2}
\rho(\lambda)\ge 0,\quad \int\rho(\lambda)d\lambda=1
\ee
We then obtain the saddle point equation
\be\label{s3}
V(\lambda)+\xi=2\int\rho(\lambda')\log(\lambda-\lambda')d\lambda',\quad
\lambda,\lambda'\in \hbox{supp}(\rho)
\ee
where the support of
the function $\rho(\lambda)$ comprises $\lambda$ such
that $\rho(\lambda)\ne 0$. It emerges in this equation
because of the first condition in (\ref{s2}). The constant $\xi$ in
equation (\ref{s3}) is just the Lagrange multiplier for the second condition
in (\ref{s2}).

However, in order to use analytic tools (the Cauchy problem),
we must investigate not eq.(\ref{s3}) but its derivative
\be\label{s4}
V'(\lambda)=2\pint{\rho(\lambda')\over\lambda-\lambda'}d\lambda',\
\hbox{or}\
V'(\lambda)=\oint{\rho(\lambda')\over\lambda-\lambda'}d\lambda'.
\ee
In order to solve this equation, we introduce the function
\be
y(\lambda)\equiv {2\over N}\sum_i{1\over\lambda_i-\lambda}+V'(\lambda)=
2\pint{\rho(\lambda')\over\lambda'-\lambda}d\lambda'+V'(\lambda)
\ee
such that its imaginary part coincides with $\rho(\lambda)$ because of
(\ref{s4}). In the large $N$ limit, it satisfies the equation
\be\label{t1}
y^2(\lambda)-V'^2(\lambda)+4\int{V'(\lambda)-V'(\lambda')\over
\lambda-\lambda'}\rho(\lambda')d\lambda'\equiv y^2(\lambda)
-V'^2(\lambda)+f_{n-1}(\lambda)=0
\ee
where $f_{n-1}(\lambda)$ is a polynomial of degree $n-1$. This means that
the general solution to (\ref{s4}) is
\be\label{t2}
y^2=V'^2(\lambda)-f_{n-1}(\lambda)=\prod_{i=1}^{2n}(\lambda-\mu_i)
\ee
This equation describes
a hyperelliptic curve of genus $n-1$. It is, however,
not arbitrary for a fixed potential $V(\lambda)$, because it
follows from (\ref{t1}) that
\be
\label{t3}
y(\lambda)-V'(\lambda)=W(\lambda)
\sim \left.{2\over\lambda}+O\left(\lambda^{-2}\right)
\right|_{\lambda\to\infty},
\ee
i.e., $\mu_i$ are not independent. Here $W(\lambda)$ is the standard
loop mean~\cite{OC}.
One more restriction comes from the
normalization condition in (\ref{s2}), and one is left with $n-1$ free
moduli. This is exactly what we need for SW system given on curve
(\ref{t2}).

A solution to eq.(\ref{s4})
is parameterized by $n-1$ moduli,
i.e., these moduli span
the moduli space of planar limits of the matrix model. The function
$\rho(\lambda)$, which is imaginary part of $y(\lambda)/(2\pi)$,
has the support on
$n$ different branching cuts.

Note that within the standard matrix model framework, there are two more
requirements that leave no moduli in solution. First of all, one can easily
see that the sign of $\rho(\lambda)$ changes when coming to the next cut.
This spoils non-negativity of $\rho(\lambda)$ and means that solution
(\ref{t2}) is not stable. This means that
the cuts with $\rho(\lambda)$ negative must
shrink to produce double points. Thus, $y(\lambda)$ becomes proportional,
besides the square root of a polynomial, to some other polynomial that has
odd numbers of zeroes between cuts and, therefore, changes sign on every next
cut.

We shall explain below that one can easily include these double points into
the general SW and Whitham framework. Moreover, they allow one to construct
more general Whitham systems.\footnote{V.Kazakov suggested to overcome
non-stability of solutions without double points via some proper analityc
continuation.}

The second requirement looks more fundamental. Namely, returning to
original equation (\ref{s3}), one has to check that the Lagrange multiplier
$\xi$ is the same for {\it every} cut (while (\ref{s4}) only
guarantees it is a constant on a cut). The difference of values of $\xi$
on two neighbour cuts is equal to \cite{Jur}
\be
\label{xi}
\xi_{i+1}-\xi_i=\int_{\mu_{2i}}^{\mu_{2i+1}}y(\lambda)d\lambda
\ee
where the integral runs from the right end of the left cut to the left end of
the right cut. This gives $n-1$ additional constraints and leaves no moduli
(there still remains a freedom in the number of cuts). We must wave this
requirement in order to make the Whitham system nontrivial. So, instead
of just matrix models, it is better to speak about matrix-model-like
solutions of Cauchy problem (\ref{s4}).
So, at the moment we just
ignore this last restriction
and work with solutions to eq.(\ref{s4}), which we call matrix model
solutions. We shall return to this point later.

\paragraph{4.}

Now we associate an SW system with the planar limit of the matrix model. The
family of genus $g$ curves is described by eq.(\ref{t2}) with the restriction
(\ref{t3}). We describe it on the complex plane $\lambda$ by $n=g+1$ cuts, Fig.1.
Besides canonically conjugated $A$- and $B$-cycles, we also use the linear
combination of $B$-cycles: $\bar B_i\equiv B_i-B_{i+1}$, $\bar B_{n-1}\equiv
B_{n-1}$. Therefore, $\bar B$-cycles encircle the nearest ends of two
neighbour cuts, while all $B$-cycles goes from a given right end of the cut
to the last, $n$-th cut. For the sake of definiteness, we order all points
$\mu_i$ in accordance with their index so that $\mu_i$ is to the right of
$\mu_j$ if $i>j$.

\begin{picture}(190,55)(0,10)
\multiput(40,40)(40,0){5}{\oval(30,10)}
\multiput(28,40)(40,0){5}{\line(1,0){24}}
\multiput(28,40)(40,0){5}{\circle*{2}}
\multiput(52,40)(40,0){5}{\circle*{2}}
\multiput(47,37)(40,0){4}{\line(0,1){6}}
\multiput(73,37)(40,0){4}{\line(0,1){6}}
\multiput(47,37)(40,0){4}{\line(1,0){5}}
\multiput(47,43)(40,0){4}{\line(1,0){5}}
\multiput(68,37)(40,0){4}{\line(1,0){5}}
\multiput(68,43)(40,0){4}{\line(1,0){5}}
\multiput(55,37)(40,0){2}{\line(1,0){10}}
\multiput(55,43)(40,0){2}{\line(1,0){10}}
\put(175,37){\line(1,0){10}}
\put(175,43){\line(1,0){10}}
\put(135,37){\line(1,0){3}}
\put(135,43){\line(1,0){3}}
\put(142,37){\line(1,0){3}}
\put(142,43){\line(1,0){3}}
\put(39,20){$A_1$}
\put(59,23){$\bar B_1$}
\put(39,50){$a_1$}
\put(79,20){$A_2$}
\put(99,23){$\bar B_2$}
\put(79,50){$a_2$}
\put(119,20){$A_3$}
\put(119,50){$a_3$}
\put(137,40){$\dots$}
\put(159,20){$A_g$}
\put(172,23){$\bar B_g=B_g$}
\put(159,50){$a_g$}
\put(199,20){$A_n$}
\put(199,50){$1-\sum a_i$}
% end of 1a
%
\end{picture}

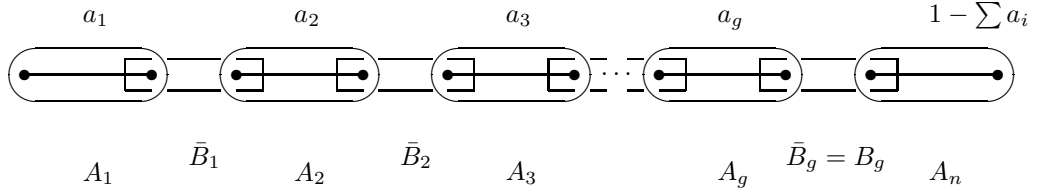
\centerline{{\bf Fig.~1.} Structure of cuts and contours.}

The SW differential is
\be\label{dS}
dS=y(\lambda)d\lambda
\ee
Its variations w.r.t. moduli is holomorphic on ${\cal C}$ (\ref{t1}) because
all moduli are hidden in the polynomial $f_{n-1}(\lambda)$:
\be
{\partial dS\over\partial \hbox{moduli}}={\partial
f_{n-1}(\lambda)\over\partial\hbox{moduli}} {d\lambda\over y}
\ee
This expression is holomorhic, because the leading coeffient of
$f_{n-1}(\lambda)$ is fixed by the normalization condition (\ref{s2}), and
the differentials $\lambda^k{d\lambda\over y}$ are holomorphic on the curve
${\cal C}$ (\ref{t1}) for $k=0,1,\dots,n-2$. Therefore, we introduce
the variables
\be
\label{vvv}
a_i=\frac12\oint_{A_i} y(\lambda)d\lambda=\Im\int_{\mu_{2i-1}}^{\mu_{2i}}
y(\lambda)d\lambda=\int_{\mu_{2i-1}}^{\mu_{2i}}\rho(\lambda)
d\lambda,\quad i=1,\dots,n-1
\ee
that have meaning of ``the occupation numbers'' (or numbers of eigenvalues)
associated with a given cut. Note also that
\be
\frac12\oint_{A_n}y(\lambda)d\lambda=\Im
\int_{\mu_{2n-1}}^{\mu_{2n}}y(\lambda)d\lambda=1-\sum_{i=1}^{n-1}
a_i
\ee
which follows from (\ref{s2}). This is exactly the condition that fixes the
leading coefficient of $f_{n-1}(\lambda)$ and leaves $n-1$ moduli. It means
that
\be\label{nc}
\frac12
\oint_{A_n}d\omega_i=
\frac12\oint_{A_n}{\partial dS\over\partial a_i}=-1\quad
\hbox{ for all }i
\ee
Now one defines the prepotential
\be
{\partial {\cal F}\over \partial a_i}=\oint_{B_i}dS
\ee
This prepotenatial is equal to logarithm of the matrix model partition
function, $\log Z_N$ in the planar limit. Indeed,
let $\cal D$ be a set of contours $A_1\cup A_2\cup\dots\cup A_g\cup
A_n$. The large $N$ partition function is\footnote{From now on,
we consider symbols $\oint$ and $res$ with
additional factors $(2\pi i)^{-1}$ so that
${\res}_0 \frac{d\xi}{\xi}
= -{\res}_\infty \frac{d\xi}{\xi} =
\oint \frac{d\xi}{\xi} = 1$.}
\be
\log Z_N=-\frac12\oint_{\cal D}y(\lambda)V(\lambda)d\lambda
+\frac{1}{4}\oint\!\oint_{\cal D\times \cal D}y(\lambda)\log(\lambda-
\lambda')y(\lambda')d\lambda\,d\lambda'
\label{b1}
\ee
We now calculate the derivative of $\log Z_N$ w.r.t. $a_i$:
\be
\frac{\partial \log Z}{\partial a_i}=
-\frac12\int_{\cal D}d\lambda
\frac{\partial y(\lambda)}{\partial a_i}
\left(V(\lambda)-\int_{\cal D}d\lambda'\log(\lambda-\lambda')y(\lambda')\right)
\label{b2}
\ee
The expression in the brackets on the rhs of~\thetag{b2} is a step
function, which is equal to $\xi_i$ on each cut $A_i$ and its values on
different cuts are (\ref{xi})
\be
\xi(\lambda)\equiv V(\lambda)-\int_{\cal D}d\lambda'\log(\lambda-
\lambda')y(\lambda')=
\left\{
\begin{array}{l}
\xi_1\equiv h_1\ \hbox{for}\ \lambda\in A_1,\cr
\xi_1+\oint_{\bar B_1}y(\lambda')d\lambda'\equiv h_2\ \hbox{for}\ \lambda\in A_2,\cr
\vdots\cr
\xi_1+\oint_{\bar B_1\cup \bar B_2\cup\dots\cup \bar
B_{g-1}}y(\lambda')d\lambda'
\equiv h_g\ \hbox{for}\ \lambda\in
A_g,\cr
\xi_1+\oint_{\bar B_1\cup \bar B_2\cup\dots\cup \bar B_{g-1}\cup
\bar B_g}y(\lambda')d\lambda'\equiv h_n\
\hbox{for}\ \lambda\in A_n.
\end{array}
\right.
\label{b3}
\ee
We therefore have
\be
\frac{\partial \log Z_N}{\partial a_i}=-\int_{\cal D}
\frac{\partial dS}{\partial a_i}h(\lambda)=
-\int_{\cal D}\omega_i h(\lambda)
=-\xi_i+\xi_n=\oint_{\bar B_i\cup \bar B_{i+1}\cup\dots\cup \bar B_g}dS
\equiv\oint_{B_i}dS
\label{b4}
\ee
and $Z_N$ in the planar limit can be, indeed, identified with the
prepotential $e^{{\cal F}}$.

\paragraph{5.}

One can learn two lessons from this fact. First of all, we can return to the
interpretation of different $\xi_i$ on different cuts within matrix model.
The standard matrix model case of equal $\xi_i$'s can be now formulated as
the set of conditions
\be
{\partial {\cal F}\over\partial a_1}=\cdots={\partial {\cal F}\over\partial
a_g}=0
\ee
These are the conditions of minimum of the matrix model partition function
w.r.t. the occupation numbers. They can be removed by introducing different
chemical potentials for different cuts.\footnote{Putting differently, one can interpret
these conditions as a criterium of stability against tunneling of eigenvalues
between different cuts \cite{David}. Stability is achieved by imposing equality
of the chemical potentials of all cuts.} We do not enter here any further
details and go instead to another lesson.

We know from studies of matrix models that their partition functions are
$\tau$-functions of some integrable hierarchies \cite{GMMMO}. What are they in
the planar limit? We have
just proved that such a partition function is an SW prepotential in
this limit. One typically associates logarithms of Whitham
$\tau$-functions with SW prepotentials
\cite{GKMMM,RG}. Therefore, we may expect that the matrix
model partition function becomes the $\tau$-function of some Whitham
hierarchy. An additional evidence for this comes from looking at the
simplest one-cut large-$N$ solution of the matrix model, when the partition
function becomes the $\tau$-function of the dispersionless Whitham
hierarchy \cite{TakTak}. Now we construct this hierarchy in very manifest terms.

First, we return to the problem of double points. Let us assume that some
of the cuts shrink, i.e.,
\be
y(\lambda)=M_{n-k}(\lambda)\sqrt{\prod_{i=1}^{2k}\left(\lambda-
\mu_i\right)}\equiv M_{n-k}(\lambda)\sqrt{g_{2k}(\lambda)}
\ee
where $M_{n-k}(\lambda)$ is a polynomial of degree $n-k$ and
$g_{2k}(\lambda)$ is a polynomial of degree $2k$. This means that one is
effectively left with a new curve
\be\label{ny}
y(\lambda)=\sqrt{g_{2k}(\lambda)}
\ee
This
curve of lower genus $k-1$ along with the differential $dS=M_{n-k}(\lambda)
y(\lambda)d\lambda$ remarkably give rise to a new SW system that depends on
$k-1$ moduli.

To see this, one needs to take into account that there still
holds eq.(\ref{t1}),
\be\label{le}
y^2(\lambda)M_{n-k}^2(\lambda)=V'^2(\lambda)-f_{n-1}(\lambda)
\equiv V'^2(\lambda)-2(V'(\lambda)W(\lambda))_+,
\ee
where we let $(\cdot)_+$ denote the polynomial part of the expression
in brackets.
Then, varying $dS$ and using (\ref{ny}), we obtain for the
{\it general\/} variation~$\delta dS$:
\be\label{v1}
\delta dS=\delta\left( M_{n-k}(\lambda)
y(\lambda)\right)d\lambda={g_{2k}(\lambda)
\delta M_{n-k}(\lambda)+
{1\over 2}M_{n-k}(\lambda)
\delta g_{2k}(\lambda) \over y}d\lambda
\ee
On the other hand, doing a variation~$\tilde\delta$ of $M_{n-k}(\lambda)
y(\lambda)$ that does not alter the potential, we obtain
from (\ref{le}) that
\be\label{v2}
\tilde\delta dS=-{1\over 2}{\tilde\delta f_{n-1}(\lambda)\over
M_{n-k}(\lambda)y(\lambda)}d\lambda.
\ee
Because this variation is a particular case of (\ref{v1}), we obtain
that zeroes of $M_{n-k}(\lambda)$ in the
denominator of (\ref{v2}) must {\it cancel},
so the maximum degree of the polynomial in
the numerator is $n-2$. The variation is then {\it holomorphic} on
curve (\ref{ny}).

This solves the problem of double points. The corresponding system with
double points (the large-$N$ limit of the matrix model) is still described by
the SW theory.

\paragraph{6.}

Let us return to the case $n=k$. This SW system is described by $n-1$ {\it
and} $n+1$ additional parameters. They are expected to be Whitham times
giving flows on the moduli of the finite gap solution. In this moduli space
there are deformations leaving the curve within the family (variations of
$f_{n-1}(\lambda)$ that does not change the potential), and those transversal
to the family. These latter are defined by the potential.
If one stays within the family of hyperelliptic
curves (\ref{t1}), there are exactly $n+1$ transversal deformations.
Therefore, in order
to have enough many deformations (Whitham times), one needs to involve
potentials of high enough degree, i.e. to deal with the construction with
double points.

In our manifest construction of the Whitham system we mainly follow
\cite{RG,MarMir} (see also \cite{Kri2}). In order to construct a Whitham
system, one needs to add to the SW data a set of punctures with local
coordinates in their vicinity. These points here are the two infinities on
the curve (\ref{ny}) and the local parameter is $\eta={1\over \lambda}$. Now
one introduces a set of meromorphic differentials $d\Omega_n$ with the poles
only at punctures (since the hyperelliptic curve (\ref{ny}) is invariant
w.r.t.  the involution $y\to -y$, from now on we just work with either of the
two infinities, see \cite{RG,MarMir}) and the behaviour
\be\label{Omega}
d\Omega_m=\left(\eta^{-m-1}+O(1)\right)d\eta,\ \ \ \eta\to 0
\ee
Then, the Whitham system is generated by a set of equations for these
differentials and the holomorphic differentials $d\omega_i$:
\be\label{We}
{\partial d\Omega_p\over\partial t_m}={\partial d\Omega_m\over\partial t_p},
\ \ \ {\partial d\Omega_m\over\partial a_i}=
{\partial d\omega_i\over\partial t_m},\ \ \
{\partial d\omega_i\over\partial a_j}={\partial d\omega_j\over\partial a_i}
\ee
These equations implies that there exists a differential $dS$
such that
\be\label{ds}
{\partial dS\over\partial a_i}=d\omega_i,\ \ \
{\partial dS\over\partial t_m}=d\Omega_m
\ee
Let us check that the differential $dS\equiv
M_{n-k}(\lambda)y(\lambda)d\lambda$ given on the curve (\ref{ny}) with the
relation for moduli (\ref{le}) really satisfies (\ref{ds}).

Indeed, we have proved the first set of relations (\ref{ds}) in the previous
paragraph. Now let us consider variations of the potential, i.e., variations
w.r.t. Whitham times $t_m$. Then, we obtain instead of (\ref{v2})
\be\label{v3}
\delta dS=-{1\over 2}{\delta \left(V'^2(\lambda)-f_{n-1}(\lambda)\right)
\over M_{n-k}(\lambda)y(\lambda)}d\lambda
\ee
while (\ref{v1}) still holds. Repeating the argument of the previous
paragraph, we conclude that the zeroes of $M_{n-k}(\lambda)$ cancel from the
denominator and, therefore, the variation may have pole only at
$\lambda=\infty$ or $\eta=0$, i.e. at the puncture. In order
to estimate this pole, one
needs to use (\ref{v1}), which implies that
$dS=M_{n-k}(\lambda)y(\lambda)d\lambda\to (V'(\lambda)
+O({1\over\lambda}))d\lambda$ and, therefore, the variation of $dS$ at large
$\lambda$ is completely determined by the variation of $V'(\lambda)$.
Parameterizing $V(\lambda)=\sum^{n+1}t_m{\lambda^{m}\over m}$ one comes
to (\ref{ds}) up to a linear combination of holomorphic differentials. One
may fix the normalization of $d\Omega_m$ that are also defined up to
a linear combination of holomorphic differentials so that eq.(\ref{ds})
would be {\it exact}. What does this normalization mean? Throughout all our
consideration we deal with $a_i$'s and $t_k$'s as independent variables.
This unambiguously defines the way $a_i$'s depend on the
coefficients of $f_{n-1}$ and is achieved merely by imposing the (obvious)
condition \cite{RG,MarMir}
\be
\label{v3_1}
\frac{\partial a_i}{\partial t_m}=\oint_{A_i}d\Omega_m=0\ \forall \ i,m,
\ee

Thus, similarly to eq.(\ref{a}) we can invariantly introduce variables $t_m$
via the relation
\be
t_m=\res_{\eta=0}\eta^m dS
\ee
and define the prepotential that depends on both $a_i$ and $t_m$ via the old
relation (\ref{F}) and the similar relation
\be
\label{v4_1}
{\partial {\cal F}\over \partial t_m}={1\over m}\res_{\eta=0}\eta^{-m}dS
\ee
One can immediately prove that such a prepotential exists \cite{RG,MarMir},
i.e., the second derivatives are symmetric, and, moreover, similarly to
Sec.~4, we find that thus defined ${\cal F}$ {\it coincides\/} with $\log
Z_N$ in the planar limit. For this, we apply the formula similar to
(\ref{b2}) with the only difference that the potential~$V(\lambda)$
itself is changed. We then obtain from (\ref{b3})
\be
\frac{\partial \log Z}{\partial t_m}=
-\frac12\int_{\cal D}d\lambda
\frac{\partial y(\lambda)}{\partial t_m}\cdot \xi(\lambda)
-\frac12\int_{\cal D}d\lambda y(\lambda) \frac{\lambda^m}{m}
=-\sum_{i=1}^g \frac{\partial a_i}{\partial t_m}(h_i-h_n)+
{1\over m}\res_{\eta=0}\eta^{-m}dS,
\label{b22}
\ee
which by virtue of (\ref{v3_1}) gives (\ref{v4_1}).
Therefore, $Z_N$ in the planar limit is the Whitham $\tau$-function,
and the whole machinery of Whitham systems works here in full strength.

\paragraph{7.}

After having constructed the large $N$ (planar) limit, next
step is to take the (double scaling) continuum limit. Namely, one has to
dwell nearby a singularity (branching point) of $\rho(\lambda)$. Say, one can
work nearby the left end of the very right, $n$-th cut $\mu_{2n-1}$
\cite{OC}. The
standard argument then is that one feels no other cuts, since they are far
away. This would mean that multi-cut solutions coincide with the one cut
solution in the continuum limit. This is, however, the case only if all
the other branching points do not come close to $\mu_{2n-1}$. Otherwise,
there exist non-trivial continuum limits \cite{Ak}. In the most non-trivial
situation, all the branching points but $\mu_{2n}$ come close to each
other\footnote{Note that the authors of the paper \cite{DV} considered the
branching point $\mu_{2n-1}$ also located far away, at some large
distance $\Lambda$. Then, they were interested in the pieces of the
prepotential that shows up a logarithmic behaviour at $\Lambda\to\infty$ plus
terms constant in $\Lambda$.
This degenerated situation better suits the $N=1$ SUSY theory needs, where
$\Lambda$ plays a role of the $\Lambda_{QCD}$ parameter \cite{CIV}. In order
to get this limit, one just suffices to rescale
$\rho(\lambda)\to{\rho(\lambda)\over\sqrt{\lambda-\Lambda}}$ with all the
$B$-cycles still encircling the point $\Lambda$. Thus, the number of cuts
becomes $n-1$, but still there is a puncture at infinity at $\Lambda$. With
properly substracted $\Lambda$-divergent pieces, it gives the same function
${\cal F}(a)$, with $a_i$ just rescaled. This gives the calculational
recipee of the paper \cite{CIV}. The only subtlety is that, in order to
properly cut-off logarithmic terms, more concretely, to reproduce the terms
$a_i^2\log\Lambda$ in the prepotential, one needs to rescale properly the matrix
model partition function by volumes of the unitary groups, see \cite{DV}.

For instance, the simplest non-trivial example of the two-cut solution
\be\label{2c}
y=\sqrt{(\lambda^2-\mu^2)(\lambda-\Lambda)}
\ee
reduces in this limit to the semi-circle distribution
$y=\sqrt{\lambda^2-\mu^2}$, but the prepotential is determined by the
integral
$
{\partial {\cal F}\over\partial a}=\int_{\mu}^{\Lambda}\sqrt{\lambda^2-\mu^2}
d\lambda
$
with $a=\int_{-\mu}^{\mu}\sqrt{\lambda^2-\mu^2}
d\lambda$. This immediately gives ${\cal F}\sim a^2\log {a\over\Lambda^2}
+...$ and coincides with the result obtained by exact calculating with
(\ref{2c}) and then bringing $\Lambda$ to infinity, although $a$ gets rescaled
by $\sqrt{\Lambda}$.
}. This is equivalent just to sending $\lambda_{2n}$ to infinity. Such a
curve describes a finite gap solution to the KdV hierarchy, moreover, the
corresponding SW system is also associated with KdV \cite{kri}. Therefore,
we expect that the matrix model partition function in this limit describes
(in leading order) the Whitham hierarchy over KdV finite gap solution.

However, it would be very instructive to construct an entire double scaling
limit in this situation, in particular, to fix proper scaling behaviours.
This means to match properly  the growth of $N$ and approaching to the singularity.
Then, one could address the problem of exact (matrix model)$\leftrightarrow$
(SW) correspondence, in particular, in integrable terms. In particular, it
would be interesting to see what is the proper deformation of SW systems in
this case.

In this respect, the problem of studying higher-genus corrections
looks very natural because the whole matrix-model-like solution
(the solution to the loop equation, see~\cite{OC}) in all genera
is completely determined by the set of data~$\{a_i,t_m\}$ and because
it was proved~\cite{AA} that spectral correlators manifest the universality
property for multi-cut solutions as well. The integrable system
that will appear in this approach must be a generalization of
a Whitham system.

At last, let us note that the construction considered in this paper is
directly extendable to other matrix models. In particular, one can consider
the model of normal matrix that has much to do with the problem of Laplacian
growth \cite{Zab}. In fact, its naive large $N$ limit describes how external
and internal moments of a domain are related. This domain is a
counterpart of the one cut. Moreover, the system is described by the
Whitham hierarchy that is the dispersionless Toda system. Now, considering
a multi-domain solution and introducing chemical potentials for different
domains, one has to get the Whitham system over a finite-gap solution to the
Toda system. This Whitham hierarchy should relate the external and internal
moments of several domains.

\bigskip

The authors are grateful to A.Gorsky, S.Gukov, V.Kazakov, A.Marshakov and
especially to A.Morozov for fruitful discussions. We are also due to
Chung-I Tan for providing us with proper references on the multi-cut
solutions.

We acknowledge the remarkable atmosphere at Mologa meeting where this
work has been started.

\end{document}